\documentclass[10pt,letterpaper]{article}
\usepackage[top=0.85in,left=2.75in,footskip=0.75in]{geometry}
\usepackage{amsmath,amssymb,mathtools}

\DeclareMathOperator{\erfc}{erfc}
\usepackage{dsfont}
\usepackage{algorithm,algpseudocode}
\usepackage{changepage}
\usepackage[utf8x]{inputenc}
\usepackage{textcomp,marvosym}
\usepackage{cite}
\usepackage{nameref,hyperref}
\usepackage[right]{lineno}
\usepackage{microtype}
\DisableLigatures[f]{encoding = *, family = * }
\usepackage{array}
\newcolumntype{+}{!{\vrule width 2pt}}
\newlength\savedwidth

\raggedright
\usepackage[aboveskip=1pt,labelfont=bf,labelsep=period,justification=raggedright,singlelinecheck=off]{caption}

\makeatletter
\renewcommand{\@biblabel}[1]{\quad#1.}
\makeatother

\date{}

\usepackage{lastpage,fancyhdr,graphicx}
\fancyheadoffset[L]{2.25in}
\fancyfootoffset[L]{2.25in}

\usepackage{lmodern}

\usepackage{comment}
\usepackage{mathrsfs}
\usepackage{empheq}

\newcommand{\bs}[1]{\boldsymbol{#1}}

\newcommand{\dif}{\mathrm{d}}

\begin{document}
	\vspace*{0.2in}
	
	\begin{flushleft}
		{\Large
			\textbf\newline{Input correlations impede suppression of chaos and learning in balanced rate networks} 
		}
		\newline
		\\
		Rainer Engelken\textsuperscript{1}*,
		Alessandro Ingrosso\textsuperscript{2},
		Ramin Khajeh\textsuperscript{1},
		Sven Goedeke\textsuperscript{3}\ddag,
		L. F. Abbott\textsuperscript{1}\ddag
		\bigskip
		\\
		\textbf{1} Zuckerman Mind, Brain, Behavior Institute, Columbia University, New York, New York, United States of America
		\textbf{2} The Abdus Salam International Centre for Theoretical Physics, Trieste, Italy
		\textbf{3} Neural Network Dynamics and Computation, Institute of Genetics, University of Bonn, Bonn, Germany
		
		\ddag These authors share senior authorship.
		
		* re2365@columbia.edu
		
	\end{flushleft}


	\section*{Abstract}
	
	Neural circuits exhibit complex activity patterns, both spontaneously and evoked by external stimuli. Information encoding and learning in neural circuits depend on how well time-varying stimuli can control spontaneous network activity. We show that in firing-rate networks in the balanced state, external control of recurrent dynamics, i.e., the suppression of internally-generated chaotic variability, strongly depends on correlations in the input. A unique feature of balanced networks is that, because common external input is dynamically canceled by recurrent feedback, it is far easier to suppress chaos with independent inputs into each neuron than through common input. To study this phenomenon we develop a non-stationary dynamic mean-field theory that determines how the activity statistics and largest Lyapunov exponent depend on frequency and amplitude of the input, recurrent coupling strength, and network size, for both common and independent input. We also show that uncorrelated inputs facilitate learning in balanced networks.
	
	
	
		
	\section*{Introduction}
	
	Neural circuits are highly interconnected, which generates complex dynamics both spontaneously and in response to incoming stimuli. Identifying mechanisms by which time-varying stimuli can control circuit dynamics is important for understanding information transmission, learning reliable input-output functions, and designing optogenetic stimulation protocols.
	
	Recurrent neural networks provide a framework for understanding the interaction between external input and internally-generated dynamics. These networks can exhibit internal rich chaotic dynamics in the absence of external input~\cite{sompolinsky_chaos_1988}. External input can suppress chaotic dynamics, thus controlling the internal state of the network~\cite{molgedey_suppressing_1992,rajan_stimulus-dependent_2010,schuecker_optimal_2018}. Control of the recurrent dynamics appears necessary for reliable task learning~\cite{sussillo_generating_2009,depasquale_full-force:_2018,ingrosso_training_2019}.
	
	Excitation and inhibition in most biological circuits is conveyed by different sets of neurons with a predominance of recurrent inhibitory feedback, a property known as 'inhibition dominance'~\cite{ozeki_inhibitory_2009,ahmadian_analysis_2013,wolf_dynamical_2014}. Moreover, neurons in local populations receive time-dependent input that is correlated across neurons and can elicit a time-dependent population response. 
	It is important to investigate how such biological features shape network dynamics, response to external inputs, and learning. A class of recurrent network models originally proposed to describe the emergence of asynchronous irregular activity is termed `balanced'~\cite{van_vreeswijk_chaos_1996,van_vreeswijk_chaotic_1998}. In these networks, large excitatory currents are dynamically canceled by strong recurrent inhibitory feedback. Firing-rate networks in the balanced state can exhibit chaotic rate fluctuations~\cite{harish_asynchronous_2015,kadmon_transition_2015}. 
How the dynamic cancellation described in binary networks~\cite{van_vreeswijk_chaos_1996,van_vreeswijk_chaotic_1998} extends to firing rate models and how it affects the suppression of chaotic activity has not yet been addressed. Previous dynamic mean-field theory (DMFT) approaches to input-driven rate networks assumed that the mean of the external inputs across neurons does not depend on time, which facilitates DMFT~\cite{molgedey_suppressing_1992,rajan_stimulus-dependent_2010,schuecker_optimal_2018}.

	It remains unclear how external inputs should be structured to suppress chaos and control the network state optimally. To address this gap, we study stimulus-induced suppression of chaos in balanced rate networks with two types of time-dependent external input. Specifically, we study time-dependent inputs that are either identical across network neurons (referred to as common input) or that vary independently for each neuron (referred to as independent input).
		
We show that it takes much stronger input modulations to suppress chaos in networks that are driven by common input, because of the cancellation of common input by strong recurrent inhibition in balanced networks. Conventional methods of dynamic mean-field theory~\cite{sompolinsky_chaos_1988, kadmon_transition_2015, harish_asynchronous_2015 , schuecker_optimal_2018} are not adequate to capture the effects of time-varying common input. Therefore, we developed a dynamic mean-field theory that is non-stationary, meaning that the order parameters are time-dependent (Materials and Methods). This novel technique accurately captures the time-dependent mean, variance, two-time autocorrelation function and the largest Lyapunov exponent of input-driven networks. Specifically, we calculate the minimum input modulation amplitude required to suppress chaos, referred to as the critical input amplitude. We examine differences between common and independent input across a wide range of frequencies of sinusoidal input modulation, gains and network sizes, using both theory and simulation. We also provide approximations at low and high input frequencies. All the analytic results match those from network simulations, provided the networks are sufficiently large.
	Our findings have important implications for learning in balanced models. We quantify how successful learning performance requires chaos suppression. As a result of residual chaos, common input that is used to suppress chaos during learning in a number of schemes~\cite{sussillo_generating_2009,depasquale_full-force:_2018, ingrosso_training_2019} meets with limited success in balanced networks unless it has a high amplitude. We show how the use of independent input resolves this problem.
	
	\section*{Results}
\begin{figure}[ !ht]
		\begin{adjustwidth}{-2.25in}{0in}
		\includegraphics{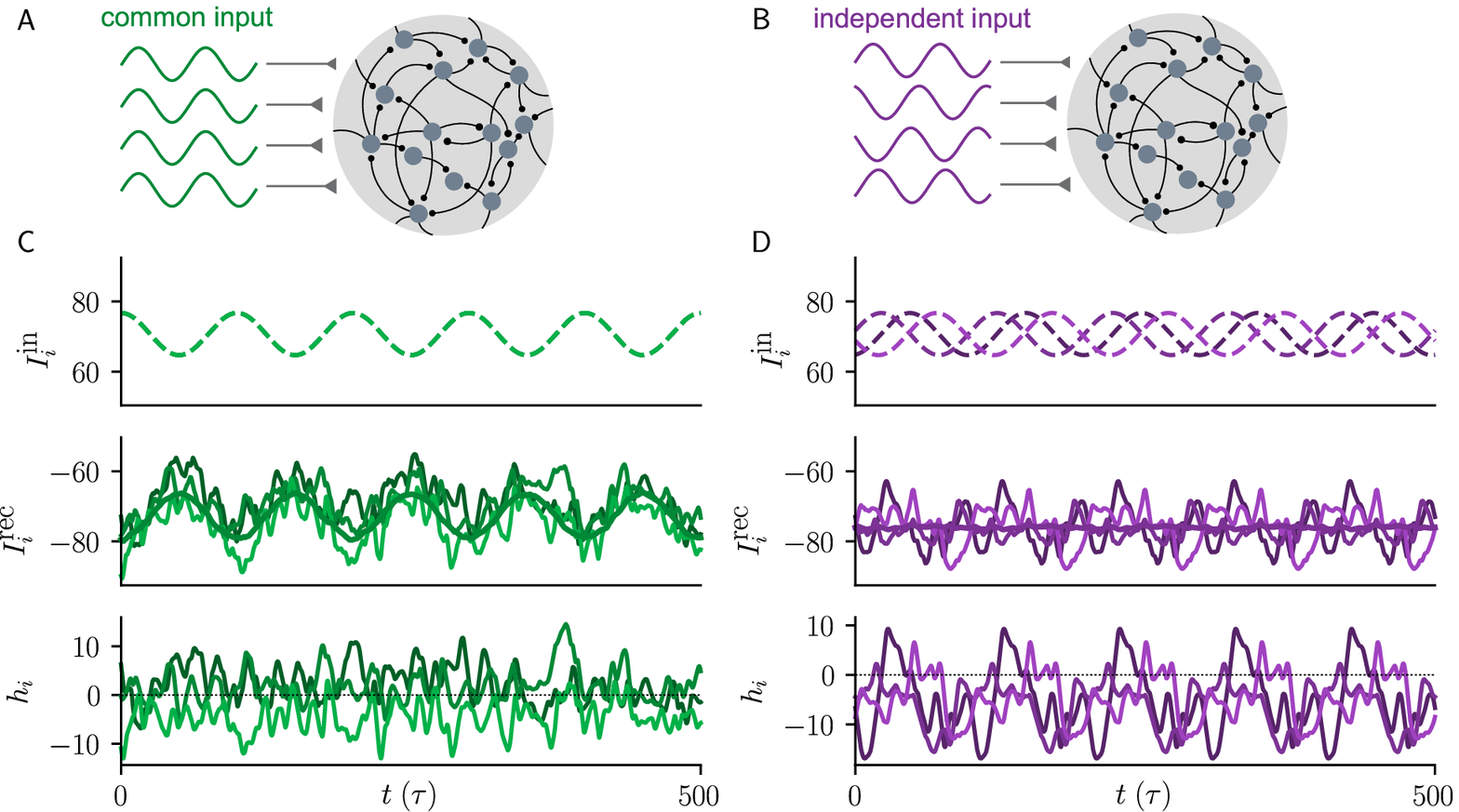}
		\vspace{0.5cm}\caption{{\bf Suppression of chaos in balanced networks with common vs independent input.}\\
			\textbf{A)}~Common input: External inputs $I_i^{\textnormal{in}} = \sqrt{N}I_0+\delta I(t)$ consist of a positive static input and a sinusoidally varying input with identical phase across neurons.
			\textbf{B)}~Independent input: External inputs $I_i^{\textnormal{in}} = \sqrt{N}I_0+\delta I_i(t)$ consist of a positive static input and a sinusoidally varying input with a random phase for each neuron.
			\textbf{C)}~External inputs (top), recurrent feedback $I_i^{\textnormal{rec}}=\sum_{j}J _{ij}\phi\left(h_{j}\right)$ and their population average (thick line) (middle), and synaptic currents (bottom) for three example neurons. Recurrent feedback has a strong time-varying component that is anticorrelated with the external input, resulting in cancellation. 
			\textbf{D)}~Same as in \textbf{C}, but for independent input. Here, no cancellation occurs and the network is entrained into a forced limit cycle. Throughout this work, green (violet) refers to common (independent) input. Model~parameters: $N=5000$, $g=2$, $f=0.01/\tau$, $I_0=J_0=1$.} 
		\label{fig1}
	\end{adjustwidth}
	\end{figure}

	We study how suppression of chaos depends on input correlations in a balanced rate network with time-dependent external input. For simplicity, we begin our analysis by studying a single inhibition-dominated population, where the recurrent inhibitory feedback dynamically balances a positive external input rather than recurrent excitation. The excitatory-inhibitory case is considered in a later section. Thus, we study a network of $N$ nonlinear rate units ('neurons') with synaptic currents $h_i$ and firing rates $\phi(h_i)$ that obey
	\begin{equation}\label{eq:mainEq}
		\tau\frac{\dif h_{i}}{\dif t}=-h_{i}+\sum_{j=1}^N J_{ij}\phi\left(h_{j}\right)+\sqrt{N}I_0+\delta I_i(t)\, ,
	\end{equation}
	with each entry of the coupling matrix $J_{ij} = -J_0/\sqrt{N} + \tilde J_{ij}$ drawn from a Gaussian distribution with negative mean $-J_0 / \sqrt{N}$ and variance $g^2 / N$, where $g$ is a gain parameter that controls the heterogeneity of weights. The transfer function $\phi$ is set to a threshold-linear function $\phi(x) = \max(x,0)$. The $1/\sqrt{N}$ scaling of the mean coupling results in strongly negative recurrent feedback that dynamically cancels the constant input term $\sqrt{N}I_0$. In addition to this constant positive term, the external input contains a time-dependent component $\delta I_i(t)$. Throughout, we distinguish between two types of time-dependent inputs, 'common' vs. 'independent'. In both cases, the time-dependence is sinusoidal, but for common input, $\delta I_i(t) = \delta I(t) = I_1 \sin(2\pi f t)$, which is identical across network neurons (Fig~\ref{fig1}A). For independent input, $\delta I_i(t) = I_1 \sin(2\pi f t+\theta_i)$ has a random phase for each neuron (Fig~\ref{fig1}B), with phase $\theta_i$ drawn independently from a uniform distribution between $0$ and $2\pi$. For simplicity, we assume that $N$ is large enough or the phases are appropriate so that we can take the average of $\delta I_i(t)$ to be zero in the independent case. The amplitude of $\delta I_i(t)$ is denoted by $I_1$, and $f$ is the input frequency.
	
	For firing-rate networks in the balanced state, suppression of chaos strongly depends on the correlations of the input (Fig~\ref{fig1}). One might expect that driving all neurons with a common input would be an effective way to suppress chaos, but input that is shared across neurons recruits strong recurrent inhibitory feedback that is anticorrelated with the common input (Fig~\ref{fig1}C). This means that the time-varying external input is dynamically canceled by recurrent feedback, leaving behind only a small fraction of the time-dependent common input for chaos suppression. In contrast, for independent input, which is randomly phase-offset across neurons, no such cancellation occurs (Fig~\ref{fig1}D), and thus weaker external input is required to suppress chaotic fluctuations in the network. 

	To understand how this discrepancy arises in the model, it is useful to rewrite Eq~\ref{eq:mainEq} by decomposing $h_{i}=m + \tilde{h}_i$ into neuron-specific and neuron-nonspecific components. For common input, this results in
	\begin{subequations}\label{eq:decompCommon}\begin{align}
			\tau\frac{\dif m}{\dif t} &= -m -\sqrt{N}J_0\nu(t)+\sqrt{N}I_0+\delta I(t)\,,\label{eq:mCommon}\\
			\tau\frac{\dif \tilde h_{i}}{\dif t} &= -\tilde h_{i}+\sum_{j}\tilde J_{ij}\phi\left(h_{j}\right)\,.\label{eq:cCommon}
	\end{align}\end{subequations}
	Here $\delta I(t)$ directly enters the expression for $m$, because it is identical across all neurons. It thus directly impacts the mean population rate $\nu(t)=\frac{1}{N}\sum_i \phi(h_i(t)) $ and recruits, through the negative recurrent mean coupling $-J_0/\sqrt{N}$, strong recurrent feedback $-\sqrt{N}J_0\nu$ that is anticorrelated with the input and cancels both the positive static input and most of the time-dependent common component of the total external input. This cancellation can be seen by rewriting Eq~\ref{eq:mCommon} as
	\begin{equation}\
			\nu(t) = \frac{I_0}{ J_0} + \frac{1}{J_0\sqrt{N}}\left(\delta I(t) -\tau \frac{\dif m}{\dif t} -m\right)\, .
	\end{equation}
This equation is commonly referred to as the 'balance equation' \cite{van_vreeswijk_chaotic_1998,kadmon_transition_2015,harish_asynchronous_2015} in the absence of time-dependent input. Note that the effect of the term $\delta I(t)$ on the population firing rate is suppressed by a factor of $1/\sqrt{N}$.
	
	With independent input, Eq~\ref{eq:mainEq} can be written as
	\begin{subequations}\label{eq:decompIndep}\begin{align}
			\tau\frac{\dif m}{\dif t} &= -m -\sqrt{N}J_0\nu(t)+\sqrt{N}I_0\,,\\
			\tau\frac{\dif \tilde h_{i}}{\dif t} &= -\tilde h_{i}+\sum_{j}\tilde J_{ij}\phi\left(h_{j}\right)+\delta I_i(t)\, .
	\end{align}\end{subequations}
	In this case, $\delta I_i(t)$ enters the equation for the fluctuations $\tilde h_i$. Thus, the strong recurrent feedback only cancels the positive static input term, $\sqrt{N}I_0$. Chaos, in this case, is suppressed through the influence of $\delta I_i(t)$ on the fluctuations $\tilde h$, similar to what happens in the non-balanced case~\cite{molgedey_suppressing_1992,rajan_stimulus-dependent_2010,schuecker_optimal_2018}.
	
	\begin{figure}[!t]		
		\includegraphics{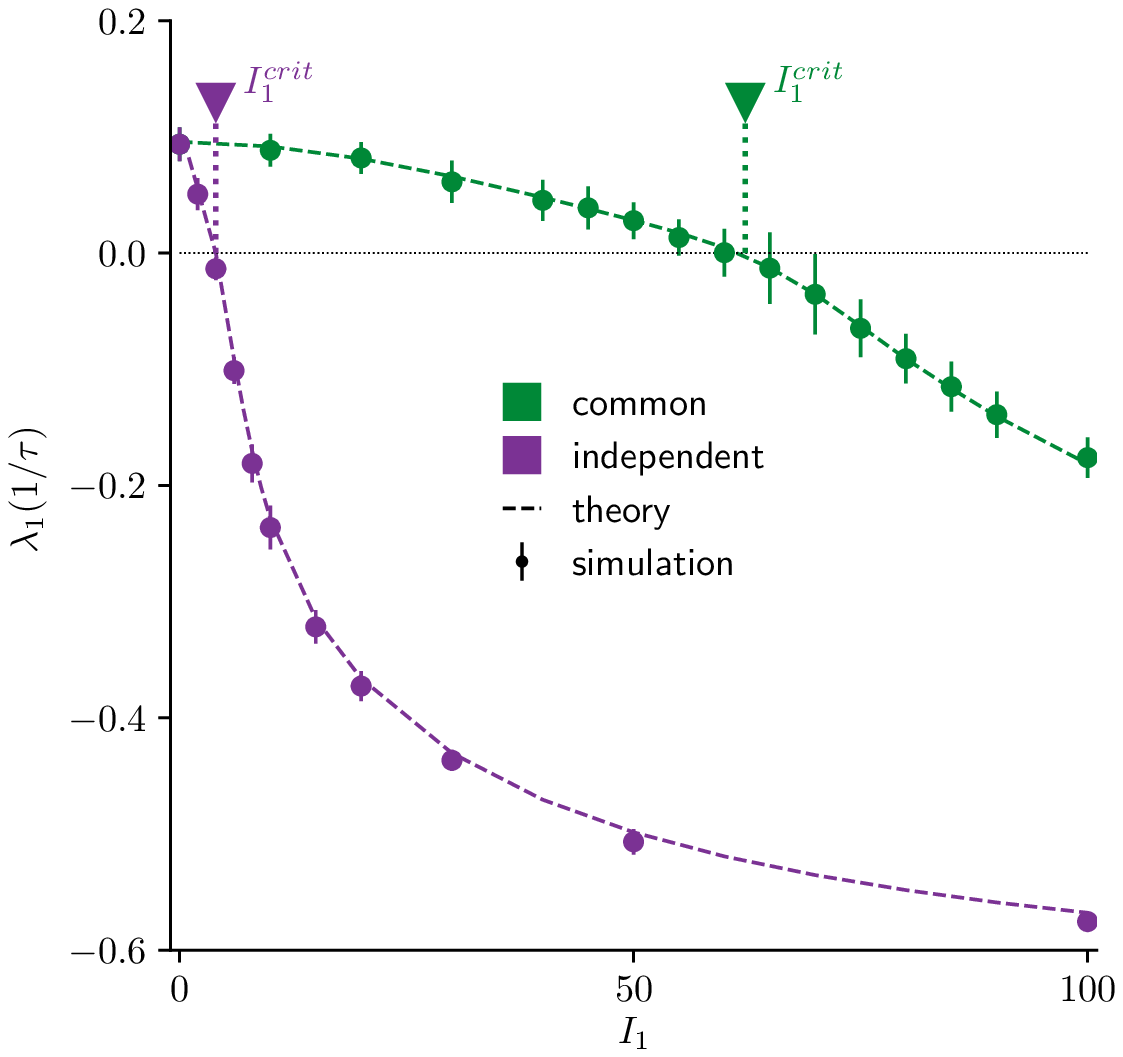}
		\caption{{\bf Largest Lyapunov exponent shows different chaos suppression for common vs. independent input.}\\
			$\lambda_1$ as a function of input modulation amplitude $I_1$ for common (green) and independent (violet) input. $I_1^{\textnormal{crit}}$ are the zero-crossings of $\lambda_1$, and thus the minimum $I_1$ required to suppress chaotic dynamics. With common input, $\lambda_1$ crosses zero at a much larger $I_1$. Dots with error bars are numerical simulations, dashed lines are largest Lyapunov exponent computed by dynamic mean-field theory. Error bars indicate $\pm 2$ std across 10 network realizations. Model parameters: $N=5000$, $g=2$, $f=0.2/\tau$, $I_0=J_0=1$.}
		\label{fig2}
	\end{figure}
	
	We quantify chaos in the network dynamics by the largest Lyapunov exponent, $\lambda_1$. This quantity measures the average exponential rate of divergence or convergence of nearby network states~\cite{engelken_lyapunov_2020} and is positive if the network is chaotic. We computed $\lambda_1$ analytically using non-stationary dynamic mean-field theory (Materials and Methods) and confirmed the finding by simulations of the full network dynamics. For both common and independent input, $\lambda_1$ is a decreasing function of the input amplitude $I_1$ and crosses zero at a critical input amplitude $I_1^{\textnormal{crit}}$ (Fig~\ref{fig2}). With common input, a much larger value of $I_1$ is required for $\lambda_1$ to become negative and thus for chaos suppression. 
	
	\begin{figure}[ !ht]	
	\begin{adjustwidth}{-2.25in}{0in}
\includegraphics{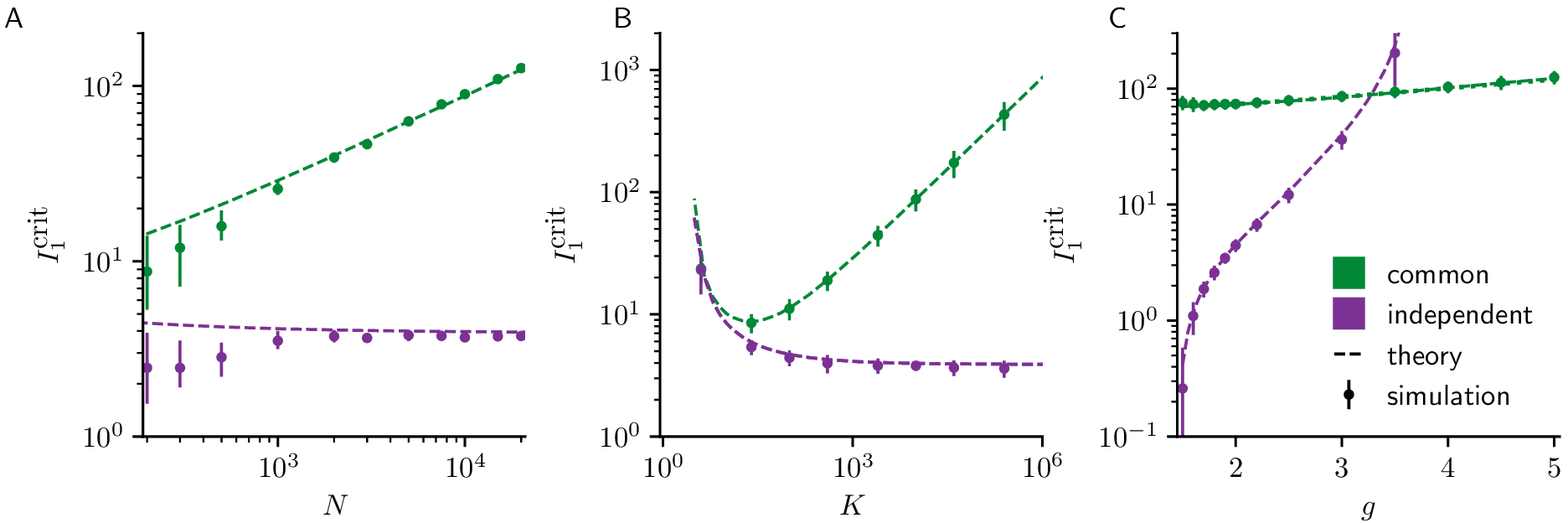}
		\vspace{0.0cm}\caption{{\bf Difference in chaos suppression increases with network size, depth of balance, and near the transition to chaos.}\\
			\textbf{A)} Dependence of $I_1^{\textnormal{crit}}$ on network size $N$. With common input, $I_1^{\textnormal{crit}}\propto \sqrt{N}$ for large $N$, but is constant for independent input. Error bars indicate interquartile range around the median.
			\textbf{B)} Dependence of $I_1^{\textnormal{crit}}$ on balance parameter $K$, which scales both $I_0$ and $J_0$. 
Results for large $K$ are the same as in A but for small $K$, the network is no longer in the strongly balanced regime, and results for common and independent input become similar. Error bars indicate $\pm 2$ std.
			\textbf{C)} Dependence of $I_1^{\textnormal{crit}}$ on gain parameter $g$ for low frequency. Close to $g_{\textnormal{crit}}$, an arbitrarily small independent input can suppress chaos; this is not the case with common input. The quasi-static approximation (dotted) and dynamic mean-field theory (dashed) results coincide. Error bars indicate $\pm 2$ std.
			Model parameters: $I_0=J_0=1$ in \textbf{A} and \textbf{C}; $g=2$, $f=0.2/\tau$ in \textbf{A} and \textbf{B}; $I_0=J_0=\sqrt{K/N}$, in \textbf{B}; $f=0.01/\tau$ in \textbf{C}, $N=5000$ in \textbf{B} and \textbf{C}.}
		\label{fig3}
	\end{adjustwidth}
	\end{figure}
	
	Next, we explore how $I_1^{\textnormal{crit}}$ varies between networks driven by common and independent input. As suggested by equations~\ref{eq:decompCommon} and ~\ref{eq:decompIndep}, the discrepancy between common and independent input grows with network size $N$. For common input, $I_1^{\textnormal{crit}}$ is proportional to $\sqrt{N}$ for large $N$, while it saturates as a function of $N$ for independent input (Fig~\ref{fig3}A). Thus an ever-increasing $I_1$ is required to suppress chaotic activity in larger networks that are driven by common input. Note that the agreement between theory and simulations is good for large $N$ (Fig~\ref{fig3}A).
	
	In balanced networks, the network size $N$ acts as a scale factor for the mean of the coupling matrix and the magnitude of the external input (Eq~\ref{eq:mainEq}). Mean-field theory describes the limit when the number of neurons goes to infinity, but it still contains $N$ as a parameter multiplying these terms. To separate these two different effects, we introduce a 'balance parameter' $K$ by scaling both $J_0$ and $I_0$ by a factor $\sqrt{K/N}$. This allows us to vary the depth of balance while still studying networks with large enough $N$ so that mean-field theory applies (Fig~\ref{fig3}B). For sufficiently large $K$, the dependence on $K$ matches that on $N$ in the unscaled model (Fig~\ref{fig3}A): for common input, $I_1^{\textnormal{crit}}$ is proportional to $\sqrt{K}$ and for independent input, $I_1^{\textnormal{crit}}$ is independent of $K$. However, the effects of independent and common input become comparable when $K\ll N$ because the model is no longer in the balanced regime.
	
	The difference in $I_1^{\textnormal{crit}}$ for common and independent input increases for decreasing $g$. With independent input, $I_1^{\textnormal{crit}}$ becomes arbitrarily small as $g$ approaches $g_{\textnormal{crit}} = \sqrt{2}$ (Fig~\ref{fig3}C). At this critical gain parameter, the network with constant external input transitions from a fixed point to chaos~\cite{kadmon_transition_2015}. At low frequency, $I_1^{\textnormal{crit}}$ remains of order $\sqrt{N}$ even near $g_{\textnormal{crit}}$ for common input (Fig~\ref{fig3}C). We note that for large values of $g$, $I_1^{\textnormal{crit}}$ for independent input becomes larger than $I_1^{\textnormal{crit}}$ for common input. The reason is that the variance of the synaptic currents $h_i(t)$ as a function of $g$ at $I_1^{\textnormal{crit}}$ grows faster for independent input than for common input as the network approaches its global instability where the dynamics diverges.
	
\begin{figure}[ !ht]
			\includegraphics{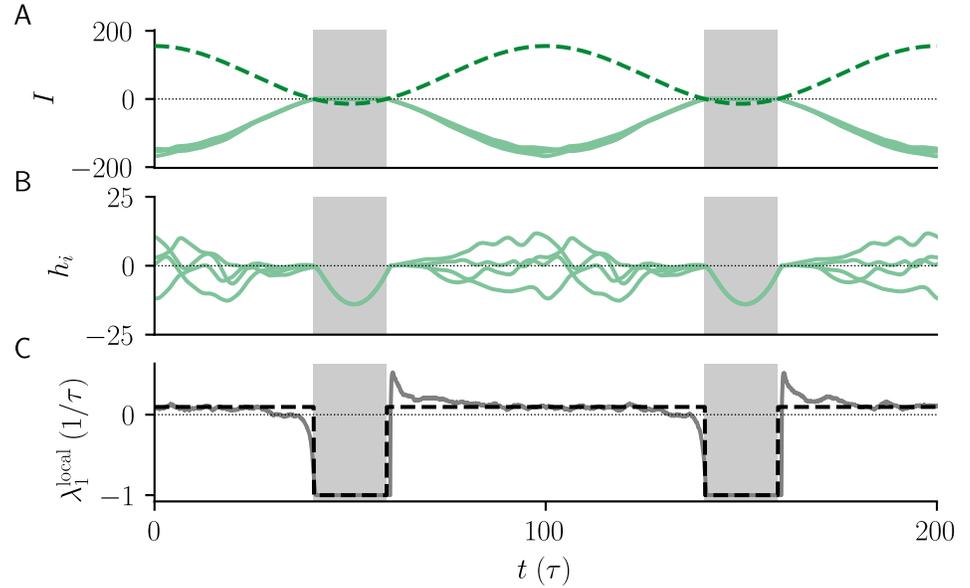}
		\caption{{\bf Mechanism of chaos suppression with slowly varying common input.}\\
			\textbf{A)}~External input $I_i^{\textnormal{in}}(t)=\sqrt{N}I_0+\delta I_i(t)$ (dashed) and recurrent input $I_i^{\textnormal{rec}}=\sum_{j}J _{ij}\phi\left(h_{j}\right)$ (solid) for three example neurons.
			\textbf{B)}~Synaptic currents $h_i$ for four example neurons. 
			\textbf{C)}~Local Lyapunov exponent from network simulation, which reflects the local exponential growth rates between nearby trajectories (solid) and Lyapunov exponents from DMFT used in quasi-static approximation (dashed). When $I_1 > \sqrt{N}I_0$, external input periodically becomes negative and silences the recurrent activity (gray bars). During these silent episodes, the network is no longer chaotic and $\lambda_1^{\textnormal{local}} = -1/\tau$. When the input is positive, dynamics remains chaotic and $\lambda_1^{\textnormal{local}} > 0$ on average. \\
			Model parameters: $N=5000$, $g=2$, $f=0.01/\tau$, $I_0=J_0=1$.}
		\label{fig4}
	\end{figure}

	An intuitive picture of chaos suppression by common sinusoidal input can be provided in the limit of low frequency, where the input varies more slowly than the intrinsic network fluctuations. In this limit, when $I_1$ exceeds the static external input $\sqrt{N}I_0$, recurrent activity is periodically silenced (Fig~\ref{fig4}A~and~B). During these silent episodes, the network dynamics is intermittently nonchaotic. On the other hand, when $g> g_{\textnormal{crit}}=\sqrt{2}$ all positive external inputs result in chaos~\cite{kadmon_transition_2015}. Thus, in a quasi-static approximation, $\lambda_1$ is given by averaging the local Lyapunov exponent $\lambda_1^{\textnormal{local}}$ across the silent and chaotic episodes, weighted by their respective durations (Fig~\ref{fig4}C; $\lambda_1^{\textnormal{local}}$ is approximated using dynamic mean-field theory). During the silent episodes, $\lambda_1^{\textnormal{local}} = -1/\tau$. In the chaotic episodes, $\lambda_1^{\textnormal{local}}$ depends on how far the network is from the transition to chaos, i.e., depends on the gain parameter $g$. As a result, $I_1^{\textnormal{crit}}$ depends on the duration of the silent episodes compared to the rest of the locally chaotic dynamics, and it grows monotonically with $g$ (Fig~\ref{fig3}C) because longer silent episodes are necessary to compensate for the stronger chaotic activity.
	 
\begin{figure}[ !ht]
	\begin{adjustwidth}{-2.25in}{0in}
		\includegraphics{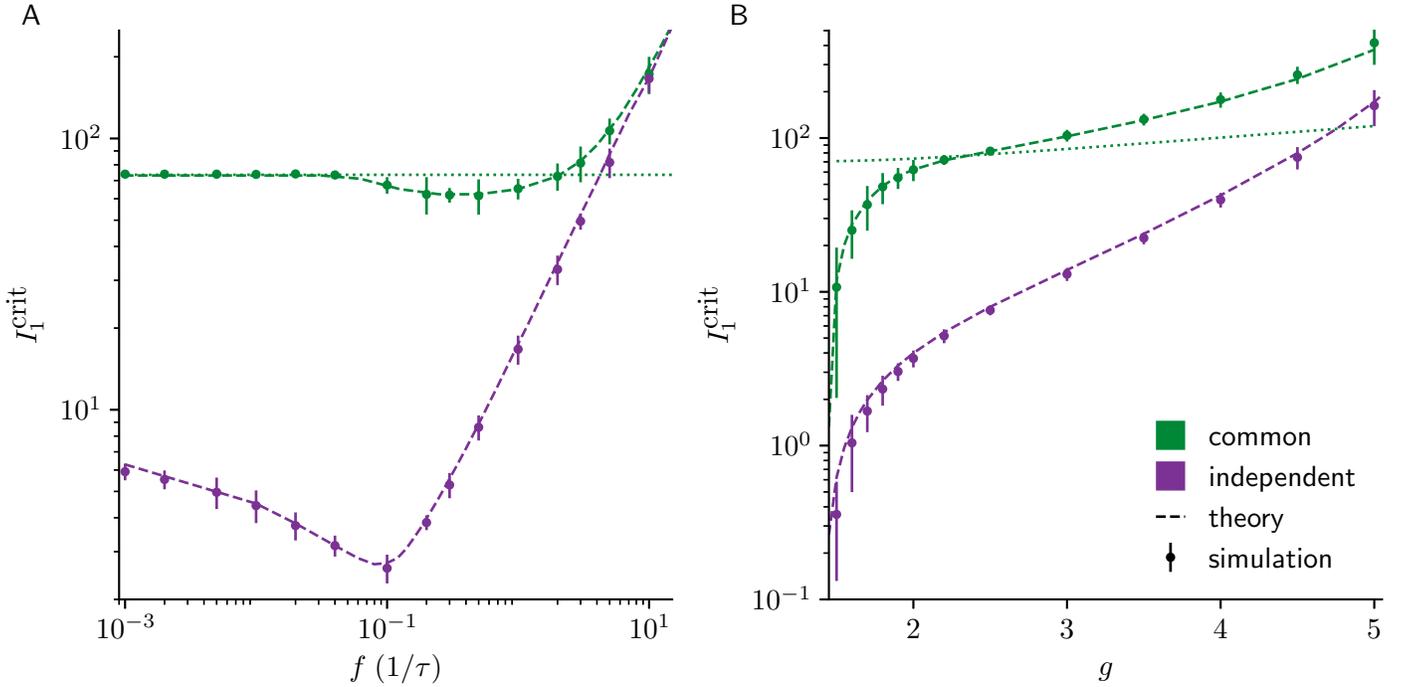}
		\caption{{\bf Dynamic mean-field theory captures frequency-dependent effects on the suppression of chaos.}
			\textbf{A)} $I_1^{\textnormal{crit}}$ as a function of input frequency. $I_1^{\textnormal{crit}}$ has a minimum that is captured by the non-stationary dynamic mean-field theory (dashed green line) but not by the quasi-static approximation (dotted green line), which does not depend on frequency. $I_1^{\textnormal{crit}}$ for independent input has a minimum. At high $f$, a low-pass filter effect of the leak term attenuates the external input for both cases, thus resulting in a linearly increasing $I_1^{\textnormal{crit}}$.
			\textbf{B)} Dependence of $I_1^{\textnormal{crit}}$ on the gain parameter $g$ for high input frequency ($f = 0.2/\tau$), showing a monotonic increase. The non-stationary mean-field theory is in good agreement with numerical simulations. For comparison, we include the result of the quasi-static approximation (dotted green line), which shows a more gradual dependence on $g$ and applies only at low frequencies. Error bars indicate $\pm 2$ std. Model parameters: $N=5000$, $g=2$, $f=0.2/\tau$, $I_0=J_0=1$.}
		\label{fig5}
	\end{adjustwidth}
	\end{figure}

	We next explore the effects of the frequency of the sinusoidal input on $I_1^{\textnormal{crit}}$ and, for both common and independent input, we observe a "resonant frequency" at which the input is most effective at suppressing chaos (Fig~\ref{fig5}A). For common input, at low frequency, $I_1^{\textnormal{crit}}$ is insensitive to the frequency and is thus well approximated by the quasi-static approximation. However, for increasing frequencies, $I_1^{\textnormal{crit}}$ exhibits a minimum in $f$, which can only be captured by non-stationary dynamic mean-field theory (Materials and Methods). For both common and independent input, when the frequency is high, low-pass filtering originating from the leak term in Eq~\ref{eq:mainEq} attenuates the effective external input modulation amplitude by a factor of $1/\sqrt{1 + 4\pi^2 f^2\tau^2}$. As a result, stronger input modulation is required to counteract the effect of this attenuation, and $I_1^{\textnormal{crit}}$ exhibits a linear increase with $f$ (Fig~\ref{fig5}A). We find that also for independent input, $I_1^{\textnormal{crit}}$ exhibits a minimum in $f$, an effect previously reported in randomly-connected networks~\cite{rajan_stimulus-dependent_2010}.
	
	We also examined the effect of coupling gain $g$ on the critical input amplitude $I_1^{\textnormal{crit}}$. For low input frequencies, a finite value $I_1^{\textnormal{crit}}$ occurred near the onset of chaos at $g=g_{\textnormal{crit}}$ (Fig~\ref{fig3}C). At a higher frequency, $f = 0.2 / \tau$, this is no longer the case (Fig~\ref{fig5}B). Close to $g_{\textnormal{crit}}$, the critical input amplitude is small for both common and independent input. 
	
	Collectively, these results demonstrate that a larger input amplitude is necessary to suppress chaotic dynamics when balanced networks are driven by common, as opposed to independent input, and that non-stationary dynamic mean-field theory successfully captures the effect in large networks.
	
	\subsection*{Results in a two population excitatory-inhibitory network}
	
\begin{figure}[ !ht]
		\includegraphics{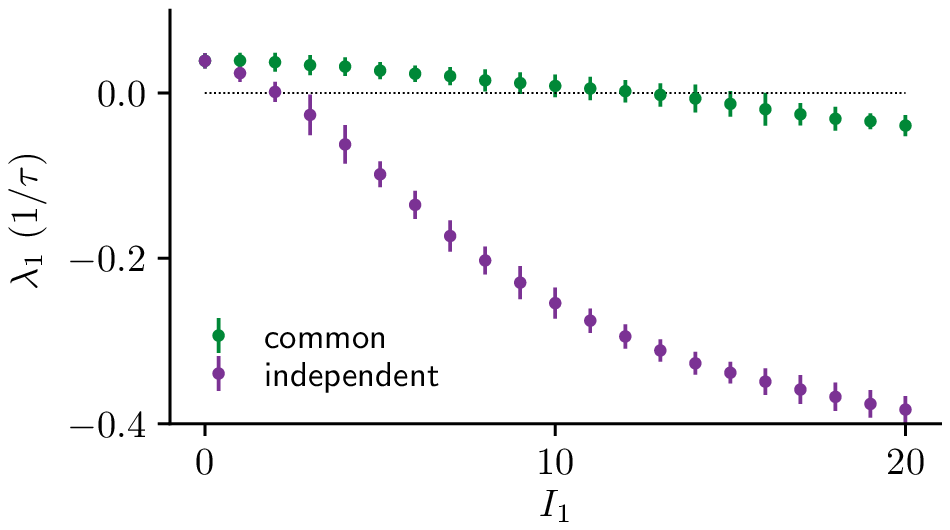}
		\caption{{\bf Difference in chaos suppression in sparsely-connected E-I network.}\\
			$\lambda_1$ as a function of $I_1$ for common and independent inputs, showing a monotonic decrease with $I_1$ and a larger zero-crossing for common input. This result is qualitatively similar to that obtained in the single population network with negative mean coupling (Fig~\ref{fig2}). Error bars indicate $\pm 2$ std. Model parameters (Parameter def. as in~\cite{kadmon_transition_2015} where $W_{I1}$ and $W_{E1}$ are the modulation amplitude of excitatory and inhibitory input): $N=7000$, $K=700$, $J_{EE}=0.88/\sqrt{K}$, $J_{EI}=-1.776/\sqrt{K}$, $J_{IE}=0.88/\sqrt{K}$, $J_{II}=-1.6/\sqrt{K}$, $W_E=0.88\sqrt{K}$, $W_I=0.704\sqrt{K}$, $W_{E1}=0.88I_1$, $W_{I1}=0.704I_1$, $f=0.2/\tau$.} 
		\label{fig6}
	\end{figure}

	The results that we report for a single population of neurons with negative mean coupling extend to a sparsely-connected two population excitatory-inhibitory network in the balanced state. We calculate the largest Lyapunov exponent $\lambda_1$ as a function of input amplitude $I_1$ and find that, consistent with our earlier observations, a much stronger input is required for common input to bring $\lambda_1$ to zero and consequently suppress the chaotic activity (Fig~\ref{fig6}).
	
	\subsection*{Common and independent input during network learning}
	
	Our results on the impact of common versus independent input have important implications for learning in recurrent networks. To address this issue, we considered a target-based approach for task learning, called full-FORCE~\cite{depasquale_full-force:_2018, ingrosso_training_2019}. In this learning procedure, a 'student network' learns a task by matching its recurrent inputs to those of a 'teacher network'. The teacher network is randomly connected and driven by the desired output to generate the target currents. The synaptic weight matrix of the student network is then trained by an online learning algorithm to autonomously generate the desired output (Materials~and~Methods).
	
	We consider a case in which the task of the student network is to autonomously generate $F^{\textnormal{out}} = \sin(2\pi f t)$. In the normal student-teacher network setup~\cite{depasquale_full-force:_2018, ingrosso_training_2019}, an input proportional to this desired output, $\delta I_i(t) = I_1 \sin(2\pi ft)$, would be injected into each unit of the teacher network. However, in a balanced network, as we have shown, this is not an efficient way to suppress chaos within the teacher network; an input of the form $I_1 \sin(2\pi ft + \theta_i)$ with varying phases would be far more effective. 
		
\begin{figure}[ !ht]
	\begin{adjustwidth}{-2.25in}{0in}
		\includegraphics{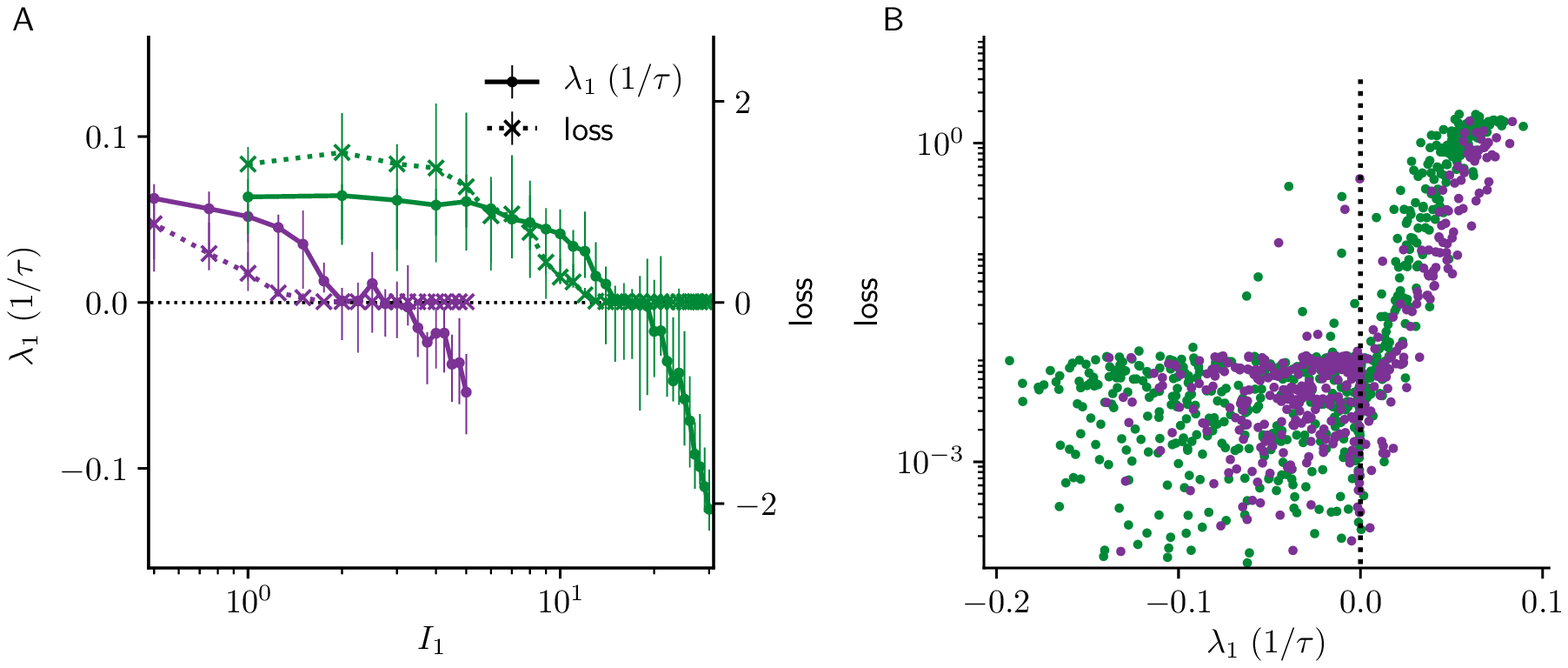}
		{\caption{{\bf Common input impedes learning in balanced networks.}\\
			\textbf{A)} Loss (test error) in the student network (solids lines) and $\lambda_{1}$ in the teacher network (dotted lines) as functions of $I_1$. The student network is trained to autonomously produce a sinusoidal output $F^{\textnormal{out}} = \sin\left(2\pi f t\right)$. Consistent with the zero-crossings of $\lambda_1$, the teacher networks driven with common input require a larger $I_1$ to achieve small test errors in the student network. Error bars indicate interquartile range around the median.
			\textbf{B)} Scatter plot of test error as a function of $\lambda_{1}$ for each network realization in \textbf{A}, with both common and independent input. When the chaos in the teacher network is not suppressed ($\lambda_1 > 0$), test error is high. Training is successful (small test error) when the targets are strong enough to suppress chaos in the teacher network. Training is terminated when loss reaches $10^{-2}$. Model parameters: $N=500$, $g=2$, $I_0=J_0=1$, $\phi(x)= \max(x,0)$ in both teacher and student networks; $f=0.2/\tau$ in the teacher network.}
		\label{fig7}}
			\end{adjustwidth}
	\end{figure}

	 We examine learning using teacher networks set up according to Eq~\ref{eq:mainEq} with each neuron $i$ driven by $I_i(t) = I_1 \sin(2\pi f + \theta_i)$. We systematically studied the influence of common input (same $\theta_i$ across the teacher network) and independent input (random $\theta_i$ across the teacher network) on learning performance in the student network. In both cases, test error drops when chaos is suppressed in the teacher network, as signaled by the zero-crossing of $\lambda_1$ (Fig~\ref{fig7}A) but a much larger value of $I_1$ is required to obtain the same test error with common input than independent input.
	
	The impact of chaos on task performance is more striking when test error is plotted against $\lambda_1$ (Fig~\ref{fig7}B), demonstrating that trained networks with small test error correspond to ones where the time-varying inputs suppress chaos in the teacher network. Interestingly, in some cases, the student network can learn to approximately reproduce the prescribed dynamics even when the teacher network is slightly in the chaotic regime (small but positive $\lambda_1$). This observation is consistent with the fact that FORCE learning and its variants can be used to build recurrent neural networks that mimic low-dimensional chaotic systems~\cite{sussillo_generating_2009}.
	
	\section*{Discussion}
	
	We investigated how correlations in the external input influence the suppression of chaos and learning in balanced networks. Stronger input modulations are required to suppress chaos when inputs are correlated across neurons. The discrepancy between common and independent input increases for large network size, deep in the balanced regime, and in the vicinity of the chaotic transition. We developed a non-stationary dynamic mean-field theory to explain the dynamic effects of time-dependent input (Materials and Methods). Lastly, we demonstrated that this discrepancy affects task learning in balanced networks.
	
	Our study is relevant in light of recent advances in optogenetics that allow for time-dependent stimulation of a selected population of neurons. Theoretical models that distinguish between different network dynamic regimes are of interest for this purpose~\cite{ahmadian_analysis_2013,ahmadian_what_2021,khajeh_sparse_2021}. 
Our work addresses this question through the spatiotemporal structure of the feedforward input. One experimental prediction of our work is that, if cortical circuits are in the balanced state, time-varying stimulation that is common across neurons will not suppress response variability as effectively as independent stimulations.

	Previous studies on suppression of chaos in rate networks were limited to independent inputs in the form of stochastic~\cite{molgedey_suppressing_1992, schuecker_optimal_2018} and sinusoidal~\cite{rajan_stimulus-dependent_2010} drive, but the networks were not balanced, and their connectivity had zero mean coupling. In these previous studies, the distribution of inputs across the population is time-independent~\cite{molgedey_suppressing_1992, rajan_stimulus-dependent_2010, schuecker_optimal_2018} and stationary dynamic mean-field theory was sufficient to describe the results. However, the treatment of common input is only possible by the non-stationary dynamic mean-field approach introduced here.

	The dynamic cancellation of time-varying input through recurrent inhibitory feedback has been previously studied in balanced networks with binary~\cite{van_vreeswijk_chaotic_1998, renart_asynchronous_2010, tetzlaff_decorrelation_2012}, and spiking neurons~\cite{darshan_canonical_2017, rosenbaum_spatial_2017}. Chaos in balanced firing-rate networks was studied previously~\cite{kadmon_transition_2015, harish_asynchronous_2015, mastrogiuseppe_intrinsically-generated_2017,schuecker_optimal_2018}, but the dynamic cancellation of correlated input and its implications on chaos suppression in rate networks were not investigated, nor were the implications for learning. It would be interesting to investigate the influence of input correlations on chaos in alternative models of the balanced state~\cite{khajeh_sparse_2021,ahmadian_analysis_2013} and rate networks with low-rank  structure~\cite{aljadeff_transition_2015,aljadeff_low-dimensional_2016,mastrogiuseppe_linking_2018,landau_macroscopic_2021}.
	

The different underlying mechanisms of chaos suppression for common and independent input we report here are not specific to periodic input modulations and threshold-linear transfer functions, which we merely chose for the sake of simplicity and analytical tractability. Networks driven by stochastic inputs, such as an Ornstein-Uhlenbeck (OU) process, exhibit a qualitatively similar discrepancy between common and independent inputs (Materials and Methods). In that case, common input corresponds to a case where all neurons receive the same realization of the OU process, with the intensity of the noise serving as the input amplitude. Moreover, a similar qualitative difference between independent and common input is expected in spiking balanced networks with sufficiently slow synaptic dynamics \cite{harish_asynchronous_2015}.

The ability to control the dynamics of recurrent networks is closely linked to the problem of learning. A target-based approach to supervised learning in recurrent networks provides a convenient framework for studying the link between chaos and trainability. This is because in these approaches, as opposed to backpropagation through time, for example, learning-induced changes in connectivity are uncoupled from the dynamics: whether chaos is suppressed in the teacher network does not depend on synaptic changes in the student network. We found that the impact of common input on chaos suppression is reflected in the learning performance: when the targets fail to suppress chaos in the teacher network, trajectories cannot be learned reliably and, as a result, the student network fails to learn the task.

	Based on our analysis, we propose two strategies to overcome this problem. One strategy is to phase-offset the target across neurons (as in independent input explored in this study) so that their population average is zero. An alternative approach is to project the target through input weights with a vanishing population average. Both solutions avoid a large time-varying mean component in the external input that would otherwise be dynamically canceled by recurrent feedback. In sum, this finding can help to harness the computational capabilities of balanced networks for learning stable trajectories. 
	
	\section*{Materials and Methods}
	
	We analyze the dynamics of Eq~\ref{eq:mainEq} with time-dependent common or independent external input. For common input, we develop a novel non-stationary dynamic mean-field theory (DMFT) yielding two-time autocorrelation functions of the activity fluctuations and the largest Lyapunov exponents. For independent input, we calculate autocorrelation functions and Lyapunov exponents using stationary DMFT~\cite{sompolinsky_chaos_1988,kadmon_transition_2015,harish_asynchronous_2015,schuecker_optimal_2018}, extending previous work~\cite{rajan_stimulus-dependent_2010}.
	
	We consider a single population of neurons with negative mean coupling, with the dynamic equation (see result section Eq~\ref{eq:mainEq}),
	\begin{align*}
		\tau\frac{\dif h_{i}}{\dif t}=-h_{i}+\sum_{j}J_{ij}\phi\left(h_{j}\right)+\sqrt{N}I_0+\delta I_i(t)\,.
	\end{align*}
	As mentioned in the main text, we decompose $J_{ij} = -J_0/\sqrt{N} + \tilde J_{ij}$, where the entries of $\tilde J_{ij}$ are i.i.d. Gaussian with variance $g^2/N$ and mean zero. 
	For convenience, we include here the decompositions of Eq~\ref{eq:mainEq} for common input,

	\begin{subequations}\begin{align}
			\tau\frac{\dif m}{\dif t} &= -m -\sqrt{N}J_0\nu(t)+\sqrt{N}I_0+\delta I(t)\,,\\
			\tau\frac{\dif \tilde h_{i}}{\dif t} &= -\tilde h_{i}+\sum_{j}\tilde J_{ij}\phi\left(h_{j}\right)
	\end{align}\end{subequations}
	(with the network-averaged (population) firing rate $\nu(t)=\frac{1}{N}\sum_{i=1}^{N}\phi\left(m(t)+\tilde{h}_{i}(t)\right)$), 
	and for independent input,
	\begin{subequations}\begin{align}
			\tau\frac{\dif m}{\dif t} &= -m -\sqrt{N}J_0\nu(t)+\sqrt{N}I_0\,,\\
			\tau\frac{\dif \tilde h_{i}}{\dif t} &= -\tilde h_{i}+\sum_{j}\tilde J_{ij}\phi\left(h_{j}\right)+\delta I_i(t)\,.
	\end{align}\end{subequations}
	After solving the non-stationary dynamic mean-field theory (DMFT) with common input, we analyze the small and large frequency limits. 
	
	\subsection*{Common input}
	
	\subsubsection*{Non-stationary dynamic mean-field theory}
	In this section, we derive a non-stationary DMFT for common input. With time-dependent common input to all units, $m(t)$ and the autocorrelation function of $\tilde h_i$ change over time. Therefore the statistics of $h_i$ is not stationary in this case in contrast to conventional DMFT approaches~\cite{sompolinsky_chaos_1988,kadmon_transition_2015,harish_asynchronous_2015,rajan_stimulus-dependent_2010,stern_dynamics_2014,schuecker_optimal_2018,muscinelli_how_2019}. 
	\begin{figure}[ !ht]
			\begin{adjustwidth}{-1.5in}{0in}
	\includegraphics{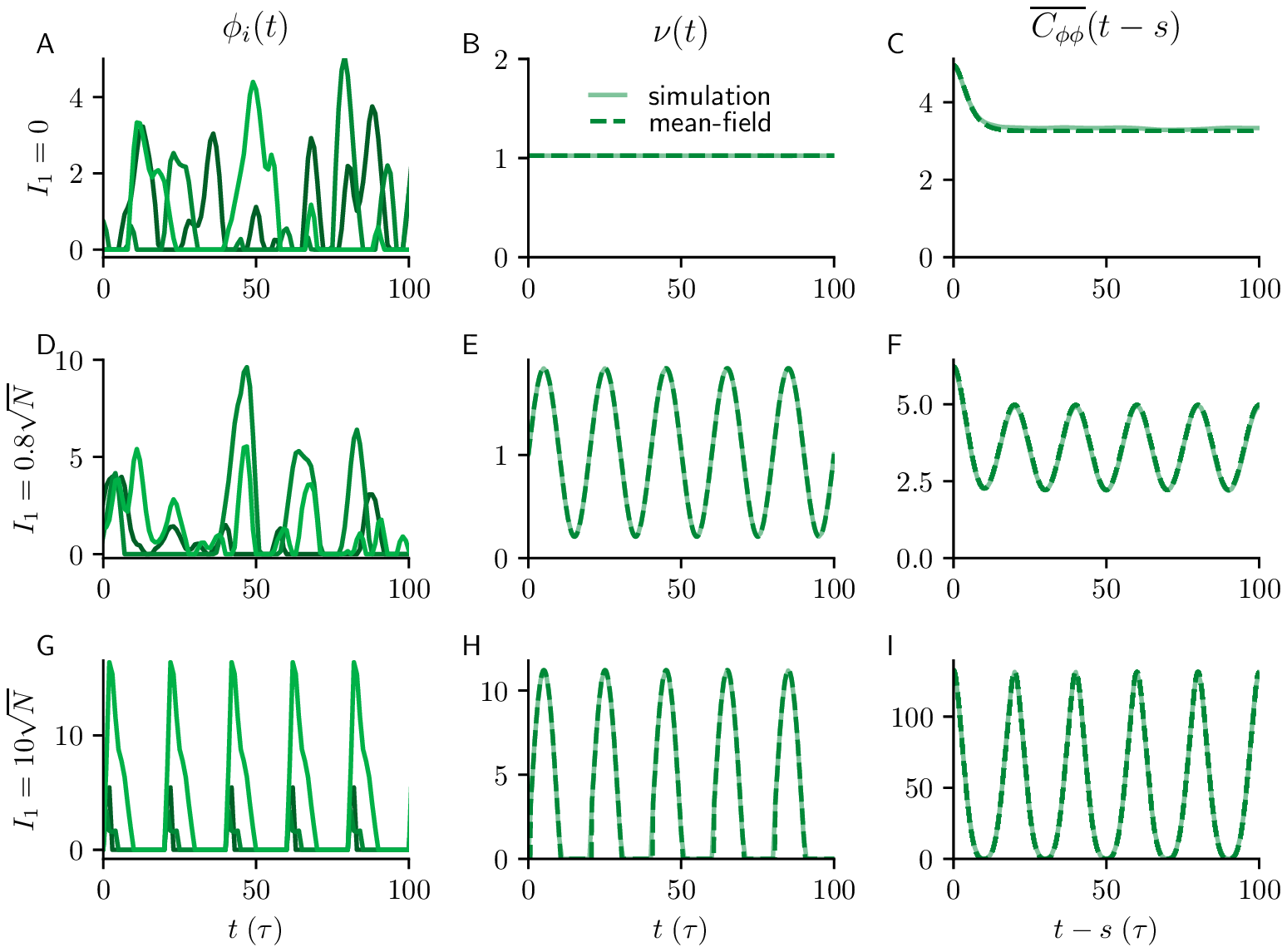}
		\caption{{\bf Activity and autocorrelations of balanced network with common input.}\\
			\textbf{A)} Firing rates $\phi_i(t)=\phi(h_i(t))$ of three example units.
			\textbf{B)}~Mean population firing rate $\nu(t)$.
			\textbf{C)} Time-averaged two-time autocorrelation function as a function of time difference with no external input ($I_1=0$).
			\textbf{D-F)} Same as \textbf{A-C} but for input amplitude of $I_1=0.8\sqrt{N}$; activity remains chaotic. \textbf{G-I} Same as \textbf{A-C} but for stronger input ($I_1=10\sqrt{N}$); activity is entrained by the external input and is no longer chaotic. Dashed lines (middle and right columns) are results of non-stationary DMFT, full lines are median across 10 network realizations. Model parameters: $N=5000$, $g=2$, $f=0.05/\tau$, $I_0=J_0=1$. }
		\label{fig8}
		\end{adjustwidth}
	\end{figure}

	\begin{figure}[!ht]
			\begin{adjustwidth}{-1.5in}{0in}
		\includegraphics{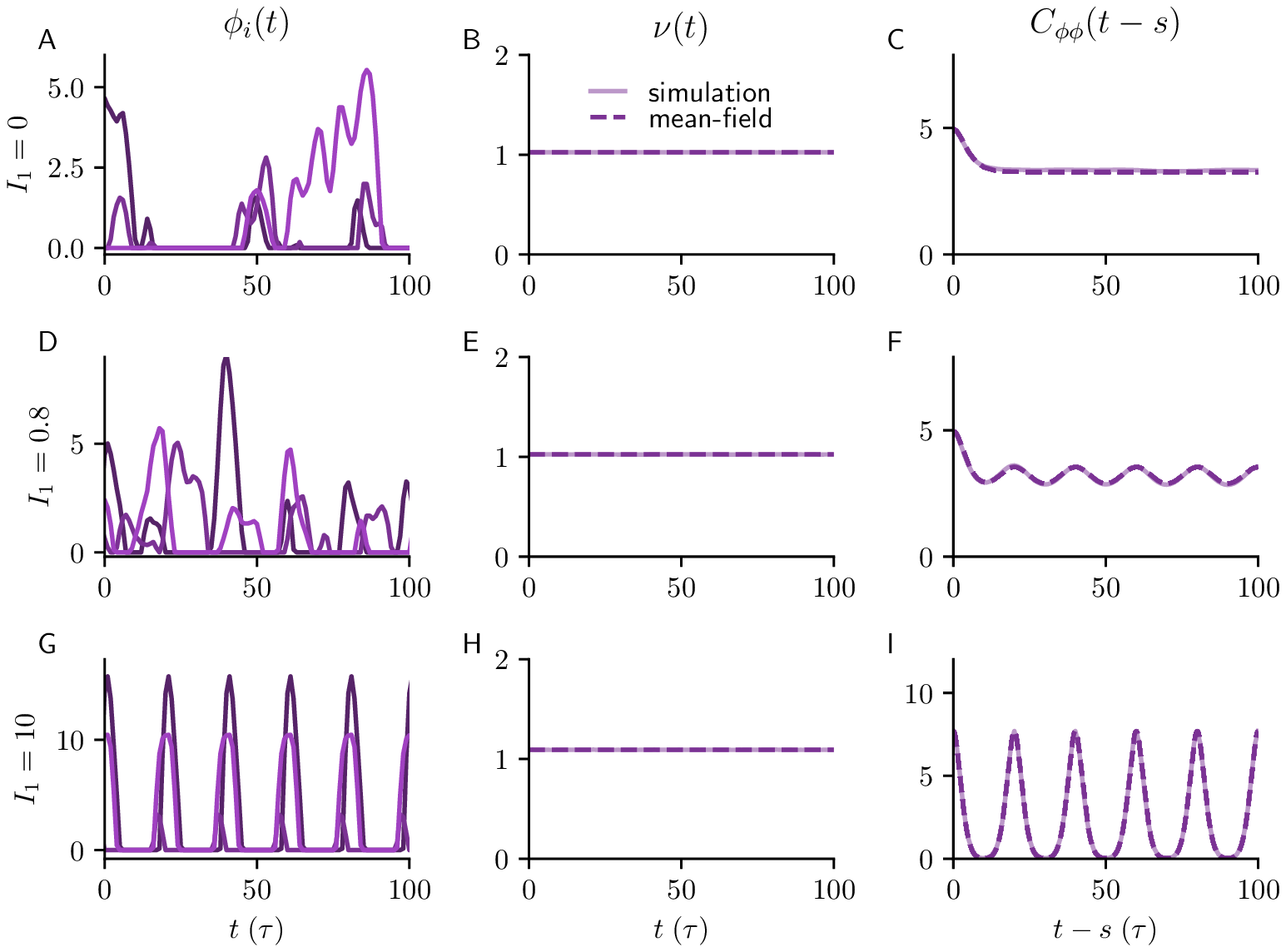}
		\caption{{\bf Activity and autocorrelations of the network with independent input.}\\
			\textbf{A)} Firing rates $\phi_i(t)=\phi(h_i(t))$ of three example units.
			\textbf{B)} Mean population firing rate $\nu(t)$.
			\textbf{C)} Autocorrelation function with no external input ($I_1=0$).
			\textbf{D)-F)} Same as \textbf{A-C} but for input amplitude of $I_1=0.8$; activity remains chaotic. \textbf{G)-I)} Same as \textbf{A-C} but for stronger input ($I_1=10$); activity is fully controlled by the external input and is no longer chaotic. Dashed lines (middle and right columns) are results of stationary DMFT, full lines are median across 10 network realizations. Model parameters: $N=5000$, $g=2$, $f=0.05/\tau$, $I_0=J_0=1$.} 
		\label{fig9}
				\end{adjustwidth}
	\end{figure}
	The basic idea of DMFT is that for large $N$, the distribution of the recurrent input for different neurons becomes Gaussian and pairwise uncorrelated, according to the central limit theorem. To this end, we characterize the distribution of the $\tilde{h}_{i}(t)$ by considering the (linear) stochastic dynamics:
	\begin{eqnarray}
		\tau \frac{\dif \tilde h}{\dif t} = -\tilde{h} + \eta(t)\,,\label{ddmft_h}
	\end{eqnarray}
	where $\eta(t)$ is a Gaussian process with mean $\langle \eta(t) \rangle = 0$ and autocorrelation
	\begin{eqnarray}
		q(t,s) = \langle \eta(t)\eta(s)\rangle = g^2 \big\langle \phi(m(t)+\tilde{h}(t))\phi(m(s)+\tilde{h}(s))\big\rangle\,.\label{ddmft_q}
	\end{eqnarray}
	Here and in the following, angular brackets denote expectation values over the distribution of the stochastic process $\tilde h(t)$, which approximates population averages in the full network. The mean-field estimate for the mean $m(t)$ therefore evolves according to Eq~\ref{eq:decompCommon}a with $\nu(t) = \langle \phi(m(t)+\tilde{h}(t)) \rangle$, the mean-field estimate of the mean population firing rate.
	
	We obtain an expression for the time evolution of the two-time autocorrelation function $c(t,s)=\left\langle \tilde h(t) \tilde h(s)\right\rangle$, which explicitly depends on two time points. Taking the temporal derivative of $c(t,s)$ with respect to $s$ and using Eq~\ref{ddmft_h}, we obtain
	\begin{equation}
		\tau\frac{\dif}{\dif s} c(t,s) =-c(t,s) + r(t,s) \label{ddmft_C}
	\end{equation}
	where $r(t,s)=\left\langle \tilde h(t) \eta(s)\right\rangle$ which we take as an auxiliary function. Taking the temporal derivative of $r(t,s)$ with respect to $t$ we arrive at an expression for the time evolution of the function $r(t,s)$:
	\begin{equation}
		\tau \frac{\dif }{\dif t} r(t,s) = -r(t,s)+ q(t,s) \label{ddmft_R2}\,,
	\end{equation}
	where $q(t,s) = g^2 \big\langle \phi(m(t)+\tilde{h}(t))\phi(m(s)+\tilde{h}(s))\big\rangle$.
	The idea to consider an auxiliary function $r$ has been proposed for a discrete-time model previously \cite{wainrib_local_2016}.
	Together, the dynamic mean-field equations for $m(t)$, $c(t,s)$ and $r(t,s)$ form a closed system of self-consistent dynamic equations and can be solved forward in time $s$ and $t$ by integrating them on a two-dimensional grid from some initial condition for $m$, $c$ and $r$. The integration requires $q(t,s)$, which can be calculated by evaluating a Gaussian double integral that depends on $c(t,s)$, $c(t,t)$, $c(s,s)$, $m(t)$ and $m(s)$. For the threshold-linear transfer function $\phi(x)=\max(x,0)$, one integral can be evaluated analytically, which allows for an efficient numerical implementation using adaptive Gauss–Kronrod integration.
	The non-stationary dynamic mean-field theory captures accurately the time-dependent mean population rate $\nu(t)$ and the two-time autocorrelation function from numerical simulations (Fig~\ref{fig8}) both in the (cyclostationary) chaotic and in the (periodic) driven stable regime. 
	
	To quantify chaos, we calculate the largest Lyapunov exponent using dynamic mean-field theory by considering the distance between two replicas of the system with identical realization of the network structure $J_{ij}$, identical external input $ \delta I(t)$, but different initial conditions~\cite{van_vreeswijk_chaotic_1998,schuecker_optimal_2018,derrida_random_1986}.
	The squared distance between the two systems can be expressed in terms of their two-time autocorrelations $c^{11}$, $c^{22}$, and the cross-correlations $c^{12}$, $c^{21}$ between them~\cite{schuecker_optimal_2018},
	\begin{eqnarray}
		d(t,s)=c^{11}(t,s)+c^{22}(t,s)-c^{12}(t,s)-c^{21}(t,s)\,.
	\end{eqnarray}
	with $c^{21}(t,s) = c^{12}(s,t)$. We next linearize the dynamics of the cross-correlation function and thereby of the squared distance around the solution that is perfectly correlated between the two replicas: $c^{12}(t,s)=c(t,s)+\epsilon\,k(t,s)\,,\:\epsilon\ll1$. This yields a linear partial differential equation for the temporal evolution of the squared distance between infinitesimal perturbations~\cite{schuecker_optimal_2018}:
	\begin{equation}
		\left(\tau\partial_{t}+1\right)\left(\tau\partial_{s}+1\right)k(t,s)=q_{\phi'\phi'}(t,s) k(t,s)\,, \label{ddmft_kProblem}
	\end{equation}
	with $d(t,t)=-2\epsilon\,k(t,t)$ and $q_{\phi'\phi'}(t,s) = g^2 \big\langle \phi'(m(t)+\tilde{h}(t))\phi'(m(s)+\tilde{h}(s))\big\rangle$. 
	
	In contrast to earlier approaches~\cite{sompolinsky_chaos_1988,kadmon_transition_2015,schuecker_optimal_2018}, where the autocorrelation was stationary, for common input, the two-time autocorrelation function is required to evaluate Eq~\ref{ddmft_kProblem}, which makes $q_{\phi'\phi'}(t,s)$ explicitly dependent on $t$ and $s$ and not only on the difference $t-s$. Eq~\ref{ddmft_kProblem}~can be solved by integrating forward on a two-dimensional grid similarly to the solution of the two-time autocorrelation function.
	
	Specifically, similar to the case of the equations for $c$ and $r$, we write equations for $k$ and an auxiliary variable $l$,
	\begin{equation}
		\tau	\frac{\dif}{\dif s} k(t,s) =- k(t,s) + l(t,s) \label{ddmft_k}\,,
	\end{equation}
	and
	\begin{equation}
		\tau	\frac{\dif}{\dif t} l(t,s) =-l(t,s)+ k(t,s) q_{\phi'\phi'}(t,s) \label{ddmft_l}\,.
	\end{equation}
	The function $q_{\phi'\phi'}(t,s)$ can be calculated by evaluating a Gaussian double integral that depends on $c(t,s)$, $c(t,t)$, $c(s,s)$, $m(t)$ and $m(s)$ which we obtained above (Eq \ref{ddmft_q}-\ref{ddmft_R2}).
	
	The largest Lyapunov exponent is given by the logarithm of the average growth rate of $k(t,t)$, discarding an initial transient:
	\[\lambda_{1}=\lim_{t\rightarrow\infty}\frac{1}{2t}\log\frac{\left|k(t,t)\right|}{\left|k(0,0)\right|}.\]
	
	\subsubsection*{Low-frequency limit}
We consider slow common input modulations $\tau\, f\ll1$. In this case, the network can be described by stationary DMFT which, for $g>\sqrt{2}$, yields chaotic dynamics for any constant positive external input~\cite{kadmon_transition_2015}. However, when $\delta I(t) < -\sqrt{N}I_0$, neurons are driven by negative input and the network becomes silent. 
During these silent episodes, because of the dissipation coming from the leak of the individual neurons, the dynamics is transiently very stable. In other words, for silenced networks, the largest Lyapunov exponent is $\lambda_1^{\textnormal{local}} = -1 /\tau$ because the Jacobian matrix of the dynamics is $ -\frac{1}{\tau}\delta_{ij}$.

The critical input amplitude $I^{\textnormal{crit}}_1$ occurs when these silent episodes on average compensate locally chaotic episodes. Since the Lyapunov exponent of the chaotic episodes is small for $g$ close to $g_{\textnormal{crit}}=\sqrt{2}$, very short silent episodes suffice to suppress chaos. Therefore, the critical input amplitude in the small-$g$ limit of the quasi-static approximation is expected to be 
\begin{equation}
	I^{\textnormal{crit}}_1=\sqrt{N}I_0.
\end{equation}
For increasing $g$, the positive input episodes become locally more chaotic, which increases $ I^{\textnormal{crit}}_1$. Thus, in the quasi-static approximation, the largest Lyapunov exponent depends on $g$ and the distribution of the time-varying input
\begin{eqnarray}
	\lambda_{1}(g,\;I_1) & = & \lim_{T\to\infty}\frac{1}{T}\int_{0}^{T}\lambda_{1}^{\textnormal{local}}(g,\sqrt{N}I_0+\delta I(t))\;\dif t\\
	& = & -\frac{1}{\tau}\int_{-\infty}^{0}p(I)\;dI+\lambda^{\textnormal{const}}_{1}(g) \int_{0}^{\infty}p(I)\;dI \,,
\end{eqnarray}
where $\lambda^{\textnormal{const}}_{1}(g)$ is the largest Lyapunov exponent for constant input~\cite{kadmon_transition_2015}, and $I$ is integrated over the probability distribution of values of $\delta I(t)+\sqrt{N}I_0$. In the last equality, we used the fact that, for constant positive external input, the Lyapunov exponent is independent of $I$ for threshold-linear transfer function because of its non-negative homogeneity. For $\delta I(t)=I_{1}\sin(2\pi ft)$ this becomes 
\begin{eqnarray}
	\lambda_{1}(g,I_{1}) & \thickapprox & -\frac{1}{\tau}\arccos\left(\frac{\sqrt{N}I_0}{I_1}\right)+\lambda_{1}^{\textnormal{const}}(g) \left(1-\arccos\left(\frac{\sqrt{N}I_0}{I_1}\right)\right)\,.
\end{eqnarray}
Solving $\lambda_{1}(g,I_{1}) = 0$ for $I_{1}$ yields
\begin{eqnarray}
	I_{1}^{\textnormal{crit}}(g) & \thickapprox & \sqrt{N}I_0\sec\left(\frac{\pi\lambda_{1}^{\textnormal{const}}(g)}{1/\tau+\lambda_{1}^{\textnormal{const}}(g)}\right)\,.
\end{eqnarray}
$\lambda_{1}^{\textnormal{}}$ is calculated analytically using dynamic mean-field theory~\cite{kadmon_transition_2015,harish_asynchronous_2015}. This is the quasi-static approximation plotted as dotted lines in Fig~\ref{fig3}C and Fig~\ref{fig5}. Note that $I_{1}^{\textnormal{crit}}$ diverges when $g$ is so large that $\lambda_{1}^{\textnormal{const}}=1/\tau$. For larger $g$, arbitrary strong slow inputs cannot suppress chaos. For $g$ close to the autonomous transition $\lim_{g\to g_{\textnormal{crit}}^{\mathrm{+}}}\lambda_{1}^{\textnormal{}}=0$, we can use the analytical approximation $\lambda_{1}^{\textnormal{const}}\left(g\right)=c\left(g-g_{\textnormal{crit}}\right)$~\cite{kadmon_transition_2015}, where $c$ is a constant of order 1.
Thus, 
\begin{eqnarray}
	I_{1}^{\textnormal{crit}}(g) & \thickapprox & \sqrt{N}I_0\left(1+\frac{\pi^{2}c^{2}}{2}\left(g-g_{\textnormal{crit}}\right)^{2}\right)\,.
\end{eqnarray}
	In the case of time-varying input generated by an Ornstein\textendash Uhlenbeck (OU) process $\frac{\dif }{\dif t} \delta I(t)=-\frac{1}{\tau_{s}}\delta I(t)+\sqrt{2D}\xi(t)$, where $\xi(t)$ is Gaussian white noise with zero mean and unit variance, a similar calculation based on $p(I)=\frac{1}{\sqrt{2\pi\tau_{s}D}}\,e^{-\frac{\left(I-\sqrt{N}I_0\right)^{2}}{2\tau_{s}D}}$ leads to
\begin{eqnarray}
	\lambda_{1}(g,\,D)
	& \thickapprox & -\frac{1}{2\tau}\erfc\left(\frac{\sqrt{N}I_0}{\sqrt{2\tau_{s}D}}\right)+\lambda_{1}^{\textnormal{const}}(g)\left(1-\frac{\erfc\left(\frac{\sqrt{N}I_0}{\sqrt{2\tau_{s}D}}\right)}{2}\right).
\end{eqnarray}
Solving by $D$ for $\lambda_{1}(g,\,D)\overset{!}{=}0$ yields
\begin{eqnarray}
	D^{\textnormal{crit}}(g) & \thickapprox & \frac{NI_0^{\,2}}{2\tau_{s}\left[\erfc^{-1}\left(\frac{2\lambda_{1}^{\textnormal{const}}(g)}{\frac{1}{\tau}+\lambda_{1}^{\textnormal{const}}(g)}\right)\right]^{2}}.
\end{eqnarray}
Again when $g$ is sufficiently large such that the largest Lyapunov exponent during the chaotic episodes reaches $\lambda_{1}^{\textnormal{const}}=\frac{1}{\tau}$, OU-input of any amplitude cannot suppress the chaos.

\subsubsection*{High-frequency limit}
For high input frequencies, the leak term in the network dynamics acts as a low-pass filter of the external input by a factor of $1/\sqrt{1 + 4\pi^2 f^2\tau^2}$. Thus, common input is low-pass filtered in Eq~\ref{eq:decompCommon}a. Analyzing the attenuation in Eq~\ref{eq:decompCommon}a, we find a linear dependence for high input frequencies,
\begin{eqnarray}
	\delta I^{\textnormal{crit}}(f)\propto\tau f\,.
\end{eqnarray}
The expected high-frequency limit is visible in Fig~\ref{fig5}A. The crossover to the linear f-dependence of $\delta I^{\textnormal{crit}}$ occurs at $ f_{c} \propto \frac{\sqrt{N}I_0}{\tau}.$ We observed such a behavior of the crossover also in numerical simulations (not shown).

\subsection*{Independent input}
\subsubsection*{Stationary dynamic mean-field theory}
In the case of independent input, we obtain autocorrelations and $m$ self-consistently analogously to~\cite{rajan_stimulus-dependent_2010} taking also a mean term into account~\cite{kadmon_transition_2015,harish_asynchronous_2015}. Moreover, we obtain the largest Lyapunov exponent analogously to previous dynamical mean-field work \cite{sompolinsky_chaos_1988,schuecker_optimal_2018}. The stationary dynamic mean-field theory captures accurately the mean population rate $\nu(t)$ and the autocorrelation function from numerical simulations (Fig~\ref{fig9}), both in the chaotic and in the driven stable regime. 

\subsubsection*{Low-frequency limit}
In the low-frequencies limit, suppression of chaos by independent input is particularly easy to understand. The network receives quenched independent input, which widens the distribution of $\tilde h$ and reduces the spectral radius and can thus suppress chaos~\cite{molgedey_suppressing_1992,schuecker_optimal_2018}. 
At values of $g$ close to the transition to chaos $g_{\textnormal{crit}}=\sqrt{2}$, only a very small input amplitude $I_1$ is necessary to suppress chaos. We find that in the zero-frequency limit
\begin{equation}
	I_1^{\textnormal{crit}}(g)=2\sqrt[4]{2}\sqrt{\pi} \frac{I_0}{J_0} \sqrt{g-g_{\textnormal{crit}}}\;.
\end{equation}
Thus, close to the transition to chaos, arbitrary small $I_1$ can suppress chaos in the independent case, while for common input, $ I_1^{\textnormal{crit}}=\sqrt{N}I_0$ in this limit. This is consistent with the results in Fig~\ref{fig3}C.
\subsubsection*{High-frequency limit}
Similar to common input, for high frequencies, the leak term in the equation for $\tilde h$ attenuates the input amplitude by a factor of $1/\sqrt{1 + 4\pi^2 f^2\tau^2}$. Thus is the high-frequency limit we expect
\begin{equation}
	I_1^{\textnormal{crit}}(f)\propto\tau f.
\end{equation}
Unlike the common input case, the scaling is not expected to depend on network size for large $N$, as the suppression of chaos is not impaired by the cancellation of the external input by recurrent feedback. This scaling is observed in Fig~\ref{fig5}A.

\subsubsection*{Quantification of chaos}
Chaotic systems are sensitive to initial conditions, and almost all infinitesimal perturbations $\epsilon\bs{u}_{0}$ of the initial condition ($\bs{h}_{0}+\epsilon\bs{u}_{0}$) grow asymptotically exponentially $|\epsilon\bs{u}_{t}|\approx \exp(\lambda_1 t)|\epsilon\bs{u}_{0}|$. The largest Lyapunov exponent $\lambda_1$ measures the average rate of exponential divergence or convergence of nearby initial conditions,
\begin{equation}
	\lambda_1(\mathbf{x}_{0})=\lim_{t\to\infty}\frac{1}{t}\lim_{\epsilon\to0}\log\frac{||\epsilon\mathbf{u}_{t}||}{||\epsilon\mathbf{u}_{0}||}\,.
\end{equation}
We calculated the largest Lyapunov exponent of the firing-rate networks in two different ways, both based on analytical expressions of the Jacobian of the dynamics~\cite{benettin_lyapunov_1980,engelken_lyapunov_2020} and with direct numerical simulations tracking the distance of two nearby trajectories. Based on the Lyapunov exponent, we computed the critical input amplitude $I_1^{\textnormal{crit}}$ using a bisection method with a relative precision of one percent.
\subsubsection*{Target-based learning}

We employ a recently developed target-based learning algorithm called full-FORCE~\cite{depasquale_full-force:_2018, ingrosso_training_2019}. The learning procedure is the following: a student network (S) learns a task by matching its total incoming currents $\eta^S_{i}\left(t\right)=\sum_{j}J^{S}_{ij}\phi\left(h^S_{j}\left(t\right)\right)$ to those of a random teacher network (T), that is driven by the desired output signal, i.e., $\eta_{i}^{T}\left(t\right)=\sum_{j}J_{ij}^{T}\phi\left(h_{j}^{T}\left(t\right)\right)+ I_i\left(t\right)$, with $I_i\left(t\right) = I_1 F^{\textnormal{out}}\left(t+\theta_i\right)$. The synaptic matrix $J_{ij}^{S}$ is trained using an online procedure so that the student network can generate the targets autonomously, $z(t) = \sum_i w_i \;\phi\left(h_i(t)\right)$, where $z(t)$ is a linear readout of the student network. Both the recurrent weights $J_{ij}^S$ and the readout weights $w_i$ are trained to produce the prescribed output signal, i.e., such that $z(t)\approx F^{\textnormal{out}}(t)$.

The incoming currents in the teacher and student network are matched via an online minimization of the following loss function for each neuron,
\begin{eqnarray}
	L_{i}=\int_{0}^{T_{tot}}\mathrm{d}\tau\Big(\eta_{i}^{T}\!\left(\tau\right)-\sum_{j=1}^{N}J^S_{ij}\phi\left(h^{S}_{j}(\tau)\right)\Big)^{2}+\alpha \sum_{j} \left(J^{S}_{ij}\right)^{2},
	\label{eq:loss_no_penalties}
\end{eqnarray}
Following~\cite{sussillo_generating_2009, depasquale_full-force:_2018}, recursive least square (RLS) is used to minimize the loss, Eq~\ref{eq:loss_no_penalties}, and to concurrently learn the readout weight vector $w_i$. We initialized both $J_{ij}^T$ and $J_{ij}^S$ as i.i.d. Gaussian matrices with average $J_0/\sqrt{N}$ and variance $g^2/N$. Euler integration was used with a time step of $\Delta t=0.01$ and $\alpha=1$.

Test error is computed over a testing period $T_{\textnormal{test}}=50T_{\textnormal{osc}}$, where $T_{\textnormal{osc}}$ is the period of the target signal, as 
	\begin{equation}
	E_{\textnormal{test}}=\frac{\int_{0}^{T_{\textnormal{test}}}\mathrm{d}\tau(\left(z\left(\tau\right)-F^{\textnormal{out}}\left(\tau\right)\right)^{2}}{\int_{0}^{T_{\textnormal{test}}}\mathrm{d}\tau\left(F^{\textnormal{out}}\left(\tau\right)\right)^{2}} \,.
	\end{equation}
For a periodic target, $F^{\textnormal{out}}$, testing is interleaved with training so that the network state $h$ is usually close to the target trajectory. In this case, a sufficiently low test error usually implies the presence of a stable limit cycle, and the periodic output is reproduced, up to a phase shift, starting from any initial condition.

\section*{Acknowledgments}
Research supported by NSF NeuroNex Award DBI-1707398, the Gatsby Charitable Foundation and the Swartz Foundation. We thank R. Darshan, J. Kadmon, A. Renart, K. Rajan, K. Miller and M. Stern for fruitful discussions.

\section*{Author Contribution}
Conceptualization: RE, AI, RK, SG, LFA.
Formal Analysis/Development of DMFT: RE, SG.
Formal Analysis/Implementation of DMFT: RE.
Software/Network simulations: RE, AI, RK.
Funding Acquisition: LFA.
Supervision: SG, LFA.
Writing – Original Draft Preparation: RE.
Writing – Review \& Editing: RE, AI, RK, SG, LFA.

\nolinenumbers

\end{document}